\documentclass[
twocolumn, preprintnumbers, amsmath, amssymb, superscriptaddress, nofootinbib, twoside]{revtex4}

\usepackage{graphicx}
\usepackage{dcolumn}
\usepackage{bm}
\usepackage{color}
\usepackage{colordvi}
\usepackage{hyperref}

\newcommand\ee{\end{equation}}
\newcommand\be{\begin{equation}}
\newcommand\eea{\end{eqnarray}}
\newcommand\bea{\begin{eqnarray}}
\newcommand{\sfrac}[2]{{\textstyle\frac{#1}{#2}}}
\newcommand\di{\partial}

\def\d{\partial}

\def\la{\langle }
\def\ra{\rangle }

\newcommand{\bg}{\begin{gather}}
\newcommand{\eg}{\end{gather}}
\newcommand{\bseq}{\begin{subequations}}
\newcommand{\eseq}{\end{subequations}}

\newcommand{\eq}[1]{(\ref{#1})}

\begin{document}


\author{Solomon Endlich}
\email{solomon@phys.columbia.edu}

\affiliation{%
Physics Department and Institute for Strings, Cosmology, and Astroparticle Physics,\\
Columbia University, New York, NY 10027, USA
}%

\author{Alberto Nicolis}
\email{nicolis@phys.columbia.edu}

\affiliation{%
Physics Department and Institute for Strings, Cosmology, and Astroparticle Physics,\\
Columbia University, New York, NY 10027, USA
}%

\author{Rafael A.~Porto}
\email{rporto@ias.edu}

\affiliation{%
Physics Department and Institute for Strings, Cosmology, and Astroparticle Physics,\\
Columbia University, New York, NY 10027, USA
}%

\affiliation{%
School of Natural Sciences, Institute for Advanced Study, \\
Einstein Drive, Princeton, NJ 08540, USA
}%


\author{Junpu Wang}
\email{junpu@phys.columbia.edu}

\affiliation{%
Physics Department and Institute for Strings, Cosmology, and Astroparticle Physics,\\
Columbia University, New York, NY 10027, USA
}%

\title{Dissipation in the effective field theory \\ 
for hydrodynamics: First order effects}

\date{\today}

\begin{abstract}
We introduce dissipative effects in the effective field theory of hydrodynamics. We do this in a model-independent fashion by coupling  the long-distance degrees of freedom explicitly kept in the effective field theory to a generic sector that ``lives in the fluid'', which corresponds physically to the microscopic constituents of the fluid. At linear order in perturbations, the symmetries, the derivative expansion, and the assumption that this microscopic sector is thermalized, allow us to characterize the leading dissipative effects at low frequencies via three parameters only, which correspond to bulk viscosity, shear viscosity, and---in the presence of a conserved charge---heat conduction. Using our methods we re-derive the Kubo relations for these transport coefficients. 

\end{abstract}

\maketitle

\section{Introduction}
Hydrodynamics has recently been recast into an effective field theory (EFT) language, with an emphasis on its internal and spacetime symmetries, their spontaneous breaking pattern,  the associated Goldstone excitations and their interactions, the derivative expansion, and in general on the systematics of the EFT program \cite{DGNR, ENRW, NS, DHNS, DHN, Nicolis}. For certain questions, this approach offers a number of advantages over the traditional one, but so far it has neglected a crucial feature of real-world hydrodynamics: the presence of dissipative effects. 
Dissipation appears in the gradient expansion of hydrodynamics as a first order correction to the perfect fluid equations, which are the continuity and the (relativistic generalization of the) Euler equations. In fact, for fluids that do not carry anomalous charges \cite{SSu}, {\em all} first order corrections are dissipative. It is then clear that for the EFT program to be useful beyond zeroth order in the derivative expansion, one has to find a way to accommodate dissipative effects.

In the standard parameterization, first order dissipative effects  are characterized by three coefficients: bulk viscosity, shear viscosity, and heat conduction.
Physically, at least for weakly coupled fluids, dissipation arises because of the fluid's microscopic constituents'  diffusion,
which tends to erase any gradients the macroscopic quantities like temperature, velocity field, etc., might have
\footnote{Bulk viscosity has a different physical origin. See e.g.~\cite{Jeon}.}.
This process effectively converts the mechanical energy carried by long wavelength perturbations, like sound waves for instance, into thermal energy. Since diffusion is essentially unobstructed when the microscopic constituents are weakly interacting, very weakly coupled fluids are, from the viewpoint of their long-distance hydrodynamical description, the {\em most dissipative} ones. This is the usual counterintuitive property of shear viscosity: at lowest order in perturbation theory, it grows linearly with the mean free path (see \cite{SS1} for a discussion about this point.) One could think that going in the opposite direction---to very {\em strong} coupling---might make diffusion and thus dissipation completely unimportant. But, apparently, this is not case. It has been conjectured that there
exists an absolute lower bound on the ratio of shear viscosity ($\eta$) to entropy density ($s$) \cite{DG, KSS},
\be \label{eta over s}
\frac{\eta}{s} \ge \frac{1}{4\pi} \; .
\ee
Interestingly, heavy ion collision data indicate that the quark-gluon plasma has an $\eta$ to $s$ ratio of the same order as the proposed bound.\\

Systems violating such a bound have been proposed---see \cite{Cremonini} for a recent review---but it is still an open question whether there exists a fundamental bound that is just somewhat lower than \eqref{eta over s}. For example, in \cite{etascaus} it has been argued that Eq. (\ref{eta over s}) can be violated for theories with gravitational duals, but still there is a (somewhat weaker) bound: $\frac{\eta}{s} \ge \frac{16}{25} \frac{1}{4\pi}$, which follows after enforcing causality in the bulk or micro-causality in the boundary CFT. 
In any case, the mere existence of such a bound still defies a purely field-theoretic justification. Therefore, part of our motivation to characterize dissipation in hydrodynamics in an EFT language, is  to derive---if it exists---a fundamental bound from sacred properties of relativistic quantum field theory. For instance, it might follow from unitarity, via dispersion relations \cite{adams,wwscat}. At this stage we make no progress in this particular direction, and keep it open for future investigation. \\

Without further ado, we now discuss how to include dissipative effects in the EFT formulation of hydrodynamics. We will paraphrase a method we learned from \cite{GR} where absorptive phenomena in black hole physics were dealt with in an EFT fashion. (See also \cite{dispin, diseft, diseft2} for generalizations, \cite{galley} for a somewhat different approach, and \cite{MSS} for an application of the same techniques in a different context.)

\section{The general idea}\label{idea}
Clearly, a  field theory with a local action is non-dissipative by construction
\footnote{The formal trick of adding an explicit time-dependence to a Lagrangian to make the energy not conserved---see e.g.~\cite{Goldstein}---might work to reproduce the desired dissipative equations of motion, but {\em (i)} is not systematic, i.e.~it is not clear what the rules of the game are, and has therefore no predictive power, and more importantly {\em (ii)} does not correspond to the physical origin of dissipation, which is that there are additional degrees of freedom that have been ignored.}. But so is Nature:
In any physical system, we call `dissipation' the transfer of energy from the degrees of freedom we are interested in (collectively denoted by $\phi$, in the following) to  others which we are not keeping track of (collectively denoted by $\chi$), either because we are not concerned about them, or because describing them is too complicated or impractical.
So,  the best way to approach dissipation from a field theory viewpoint---at least conceptually---is to keep in mind that these additional degrees of freedom should also appear in the action of the system. That is, if we were to write the full action for $\phi$ and $\chi$, we would have
\be \label{S+S+S}
S[\phi, \chi] = S_0[\phi] + S_\chi[\chi] + S_{\rm int}[\phi, \chi] \; .
\ee 
$S_0$ is the action we would write for $\phi$ alone, if we forgot about $\chi$. $S_\chi$ governs the dynamics of $\chi$. $S_{\rm int}$ couples the two sectors, and is responsible for exchanging energy between them. If we now compute observables involving our $\phi$ only, we can detect `dissipative' effects---corresponding to exciting the $\chi$ degrees of freedom---which cannot be reproduced by using $S_0$ alone.
For instance, the $S$-matrix restricted to the $\phi$-sector is non-unitary whenever producing $\chi$-excitations  is energetically allowed.

In the particular case we are interested in, $\chi$ stands for the degrees of freedom of the microscopic constituents making up the fluid. For instance, for a weakly coupled, non-relativistic fluid made up of massive point-particles, $\chi$ stands for the positions of these particles. On the other hand, $\phi$ stands for the collective degrees of
freedom, like sound waves for instance, which are those explicitly kept by the hydrodynamical description
\footnote{Strictly speaking, to avoid double counting, one should remove from the $\chi$'s the combinations of the individual particle positions that make up the $\phi$'s.}.
Notice that at all times we are dealing with {\it one and the same} fluid, and its microscopic constituents. The splitting in Eq. (\ref{S+S+S}) is one of the key features of the EFT formalism and the (emergent) dynamics in the long-wavelength limit. Hydrodynamics is about the dynamics of $\phi$.

To illustrate the general idea, in this section we will not commit to the hydrodynamical case, nor  will we go into many details. Rather, we will keep the discussion as general and as schematic as possible. We will only assume that the interaction Lagrangian $S_{\rm int}$ can be treated as a small perturbation. If this is not the case---if the two sectors are strongly coupled to each other---then it is not even clear how to talk separately of the $\phi$-sector and of the $\chi$-sector. In other words, we are assuming that as a first approximation, one {\em can} neglect the $\chi$'s when talking about the $\phi$'s. For hydrodynamics, as will see, this will be guaranteed by the symmetries: at low frequencies and momenta, {\em all} the interactions of the $\phi$'s become negligible, including those with the $\chi$'s. Notice that we are not assuming anything about interactions {\em within} the $\chi$ sector: they can be arbitrarily strong.

Now, the crucial question is how to make use of expression \eq{S+S+S}, without actually specifying what the $\chi$'s and their dynamics really are. The idea is to make  the dependence of the interaction piece $S_{\rm int}$ on $\phi$ explicit, while keeping that on $\chi$ implicit. Schematically:
\be \label{Sint}
S_{\rm int} = \int \! d^4x \sum_{n,m} \di^n \phi^m (x) \, {\cal O}_{n,m}(x) \; .
\ee
The ${\cal O}$'s are `composite operators' of the $\chi$-sector---local combinations of the $\chi$'s and their derivatives. As usual, one expects all couplings allowed by symmetry to appear in the action. So, in particular, the ${\cal O}$'s should carry spacetime and possibly internal indices in order to make the combinations appearing in $S_{\rm int}$ invariant under all the symmetries that act on the $\phi$'s. Apart from symmetry, as usual in EFT, the other organizational principle in the infinite series \eqref{Sint} is the derivative expansion: terms with fewer derivatives acting on the long distance/low energy degrees of freedom ($\phi$) matter the most at low energies and momenta.

Now, in any observable that  involves measuring the $\phi$'s only---like for instance a $\la \phi \phi \cdots \phi \ra$ correlation function, or a 
$\phi\phi \to \phi\phi$ scattering amplitude---all effects due to the presence of the $\chi$'s, dissipative or otherwise, are ``mediated'' by the correlation functions of these ${\cal O}$ composite operators. As an example, consider a coupling (linear in $\phi$) between the two sectors of the form
\be
\label{linear coupling}
S_{\rm int} = \lambda \int \! d^4 x \, \phi \, {\cal O} \; ,
\ee
where $\lambda$ is a small coupling constant. For instance, suppose that we are interested in computing the $T$-ordered two-point function of $\phi$ in the standard vacuum (i.e. the vacuum for both the $\phi$ sector and the $\chi$ sector, for as we will see in a moment, computing this same correlator with a non-vacuum dissipative $\chi$ sector will necessarily complicate the story).
This two-point function will receive contributions from $S_0$ and from $S_{\rm int}$. We can compute the latter contribution in perturbation theory for $\lambda$. For instance, if $\phi$'s only interaction is that contained in $S_{\rm int}$ above, this would correspond to the Feynman diagram series of fig.~1. In that case, neglecting combinatoric factors, powers of $i$, and momentum-conserving delta-functions, we would have schematically
\begin{align}
\label{Dyson series}
\langle \phi(p)  \phi(-p) \rangle & = \langle \phi(p) \phi (-p)\rangle _0  \\
& + \lambda^2 \langle \phi (p)\phi(-p) \rangle _0 {}^2 \, \langle {\cal O} (p){\cal O}(-p) \rangle _0  \nonumber \\
& + \lambda^4 \langle \phi(p) \phi(-p) \rangle _0 {}^3\, \langle {\cal O} (p) {\cal O} (-p) \rangle _0 {}^2+ \dots  \nonumber \; ,
\end{align}
where $T$-ordering is understood, and the subscript zeroes denote that those two-point functions are to be computed at zeroth order in $\lambda$, that is, in the absence of any interactions between $\phi$ and $\chi$. Once $\langle \phi \phi \rangle_0$ and $\langle {\cal O}  {\cal O} \rangle_0$ are known, the full $\langle \phi \phi \rangle$ can be computed at any order in $\lambda$, without any further explicit reference to the $\chi$ dynamics. 
This is analogous to the standard Feynman-diagram expansion for a perturbative QFT, which involves the free propagators only. Here the correlators on the r.h.s.~are not the free ones---they are those determined by $S_0$ (for $\phi$) and by $S_{\chi}$ (for ${\cal O}$) separately.
In a more general case, where $\phi$ has non-trivial self-interactions and couples to the $\mathcal{O}$'s in a more general way, the right-hand side looks more complicated because it involves higher-point correlation functions of $\phi$ and ${\cal O}$ as well. However, all the correlators are still evaluated at zero coupling ($\lambda$) between the two sectors.

\begin{figure} 
  \centering
      \includegraphics[width=0.45\textwidth]{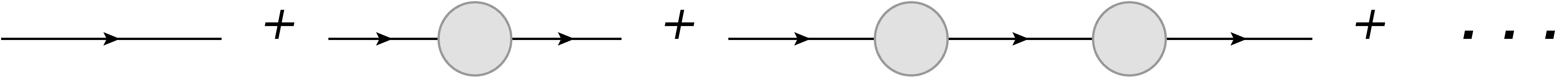}
  \caption{\em Feynman diagram representation of eq.~(\ref{Dyson series}): the solid lines represent the $\phi$ propagators and the gray circles the two-point function of $\mathcal{O}$.}
\end{figure}

As hinted at before, this simple picture gets slightly more complicated for  correlation functions in more general states and in particular, thermal states. As we will discuss at some length in the next section, we will be interested in a thermalized $\chi$ sector. Its  real-time correlation functions and the associated perturbative expansion then have to be handled via the so-called In-In, or Schwinger-Keldysh, formalism (for extensive reviews, see e.g.~\cite{Jordan, Maciejko, CH}). This entails a doubling of the fields in the path-integral, $\phi \to \phi_\pm$, $\chi \to \chi_\pm$, which  complicates somewhat the systematics of the Feynman-diagram expansion. However, for what we are interested in, we can instead consider the effective (linearized) equations of motion for the expectation value of $\phi$ that we get by ``integrating-out" the $\chi$ sector via In-In path integrals, which is essentially an In-In generalization of the quantum effective action formalism that is appropriate for systems described by a density matrix.

%

Following the notation of \cite{Jordan} and utilizing the simple coupling given by (\ref{linear coupling}), the In-In generating functional for the correlation functions of $\phi$ is given schematically by
\begin{align}
e^{iW[J_+,J_{-}]} &=\text{const} \times \int \mathcal{D} \phi_{\pm} \mathcal{D} \chi_{\pm}  \\
& e^{i(\pm S_\phi[\phi_{\pm}]\pm J_{\pm}\phi_{\pm}\pm S_\chi[\chi_{\pm}]\pm \lambda\phi_{\pm}  \mathcal{O}_{\pm})} \; ,
\end{align}
where the functional integral over $\chi_+$ and $\chi_-$ is understood to include a (thermal) density matrix $\rho(\chi_0^+, \chi_0^-)$ for the initial conditions, which are also integrated over  \cite{Maciejko, CH}. As we  will see in a second, we will not need to be explicit about this.

Let's assume that $\left< \mathcal{O}\right>=0$ and confine ourselves to quadratic order in the $\phi_{\pm}$ fields. Noticing that, from the viewpoint of the $\chi$ sector, the $\phi_\pm$ fields act as external sources for the  operators ${\cal O}_\pm$, we can formally perform the functional integration over $\chi_+$ and $\chi_-$ and obtain
\be
\label{quadratic +- path integral}
e^{iW[J_+,J_{-}]} =\text{const} \times \int \mathcal{D} \phi_{\pm}  e^{i(\pm S_2[\phi_{\pm}]\pm J_{\pm}\phi_{\pm})} e^{\frac{i \lambda^2}{2}{\phi}^a \mathcal{G}_{\mathcal{O}}^{ab}{\phi}^b} \; ,
\ee
where $S_2$ is the quadratic action for $\phi$,  ${\phi}^a \equiv (\phi_+, - \phi_-)$, and $\mathcal{G}_{\mathcal{O}}$ is a matrix of ${\cal O} {\cal O}$ correlators \cite{Jordan}:
\be
\mathcal{G}_{\mathcal{O}}(x_1,x_2)=\left( \begin{array}{cc}
\left<T\mathcal{O}(x_1)\mathcal{O}(x_2) \right> & \left<\mathcal{O}(x_1)\mathcal{O}(x_2) \right>  \\
\left<\mathcal{O}(x_2)\mathcal{O}(x_1) \right>  & \left<\mathcal{O}(x_1)\mathcal{O}(x_2) T \right>  \end{array} \right)
\ee
(the $T$ to the right of a sequence of operators implies anti-time ordering.) These correlators have to be understood as traces involving the density matrix that is appropriate for the $\chi$ sector.

The In-In effective action $\Gamma \left[{\phi}_+,{\phi}_- \right]$ is then just the Legendre transform of the In-In generating functional $W[J^+, J^-]$, from which the effective equations of motion for $\langle \phi \rangle$ follow simply as \cite{Jordan}
\be
\left . \frac{\delta \Gamma}{\delta {\phi}_+(x)}\right|_{{\phi}_+={\phi}_-= \langle \phi \rangle}=0 \;,
\ee
However, since we are working at quadratic order in $\phi_\pm$, the effective action $\Gamma$ is just whatever appears at the exponent in the path integral \eqref{quadratic +- path integral} after having set $J^\pm$ to zero:
\be
\Gamma_2[{\phi}_+, {\phi}_-] = S_2[{\phi}_+] -S_2[\phi_-] +\sfrac{\lambda^2}{2}{\phi} ^a \mathcal{G}_{\mathcal{O}}^{ab}{\phi}^b  \; ,
\ee
where two convolutions are understood for the last term.
We thus get that the linear equation of motion for the expectation value of $\phi$---which, to keep the notation light, we also call $\phi$---is simply
\be
\frac{\delta S_2}{\delta \phi} + i \lambda^2 \, \langle {\cal O \cal O} \rangle_R * \phi = 0 \; ,
\ee
where the second term involves precisely the {\em retarded} two-point function of ${\cal O}$:
\be
\langle {\cal O}(x_1) {\cal O}(x_2) \rangle_R \equiv \theta(t_1-t_2) \langle \big[ {\cal O}(x_1), { \cal O}(x_2)  \big] \rangle \; .
\ee
Note that the above conforms to the expectations of the usual ``linear-response theory'' result. What's nice about the In-In formalism is that it allows one to generalize such a result to all orders in perturbation theory in a systematic fashion.

Keeping  these qualifications in mind, and coming back to the main message of this section:
For generic $S_{\rm int}$, in order to compute observables that involve the $\phi$'s only---and in particular the time-evolution of $\left<\phi(x)\right>$---we need not be explicit about the dynamics of the $\chi$'s. We `only' need the $n$-point correlation functions of the operators the $\phi$'s couple to. Of course, knowing all such correlators is essentially equivalent
to having solved the theory defined by $S_\chi$, which, as we stressed, can be arbitrarily complicated, strongly coupled, or simply unknown. Fortunately, in our particular case of hydrodynamics, we are interested in such correlators at very low frequencies and very long distances only. Moreover, we can assume that the $\chi$ sector---whatever it is---is in a state of thermal equilbrium. As we will see, this allows us to parameterize the leading low-frequency, long-distance behavior of the relevant correlators by three coefficients only.

\section{Low frequency, long distance behavior of correlators}\label{Low_frequency}

Consider the two-point function for a generic operator in the $\chi$ sector, $\langle {\cal O}(\vec x, t) {\cal O}(\vec x \,' , t')\rangle$. If, in the absence of external perturbations---due for instance to our $\phi$'s---, the $\chi$ sector is thermalized, then the average $\langle \dots \rangle$ has to be interpreted as a thermal trace with a density matrix $\rho \propto e^{-\beta H}$, or, in the presence of a conserved charge, $\rho \propto e^{-\beta (H-\mu Q)}$. (We  will use a quantum mechanical language, but everything we say applies straightforwardly to classical statistical systems as well.) Now, we will assume that we identified correctly all the degrees of freedom that can propagate at long distances and for long times--we called them $\phi$--and that we constructed the most general EFT for them, encoded by $S_0[\phi]$. We will be more explicit in the next section, but for the moment, it suffices to say that these $\phi$'s correspond to the degrees of freedom traditionally associated with hydrodynamics: long-wavelength fluctuations in the energy density, in the velocity field, in the charge density, etc. Following the traditional language, we have `hydrodynamic modes'---i.e. physical variables with non-trivial long-range, late time correlators---for each conserved quantity: energy, momentum, charge. It is usually believed that thermal equilibrium erases all other information that is not associated with conserved charges. In particular, it is usually believed that in a thermal system correlators for quantities that are not densities for conserved charges decay rapidly, faster than any power, at very large distances and at very late times---roughly speaking, at distances and times larger than the mean free path and the mean free time, respectively. 

Following this intuition, we will assume that the $\chi$-sector only features such rapidly decaying correlators.
As we will see, this does not imply that it does not feature gapless excitations. Indeed: if there were no gapless $\chi$-excitations, it would not be possible for very low frequency $\phi$ fields to transfer any energy to the $\chi$ sector.  That is: at  frequencies lower than the gap, there would be no dissipation whatsoever. Now, if an $\langle {\cal O} {\cal O} \rangle$ correlator decays faster than any power at large space- and time-separations, then its Fourier transform
\be
\label{FourierTransform}
G(\omega, \vec k \,) \equiv \int \! d^3 x dt \, e^{i(\omega t - \vec k \cdot \vec x)} \, \langle {\cal O}(\vec x,t ) {\cal O}(0) \rangle \; ,
\ee
is differentiable for real $\omega$ and $\vec k$---infinitely many times---at $\omega=\vec k =0$. In particular, it admits a Taylor expansion in powers of $\omega$ and $\vec k$ about the origin. This means that at very  low frequencies and momenta, we can parameterize our two-point function by just a few numbers---the coefficients of the leading terms in such a Taylor expansion. 

To develop some physical intuition, it is useful to rephrase the above statement in terms of the spectral density for the operator ${\cal O}$.
So far we have been cavalier about the ordering of operators inside the two-point function. As pointed out in the last section, we will be mostly interested in  the retarded two-point function, 
\be
G_R(\vec x, t) \equiv \theta(t) \langle [{\cal O}(\vec x,t), {\cal O}(0)] \rangle \; ,
\ee
which describes the causal response of the system to external disturbances, in the sense that adding a term $\int d^3 x J {\cal O}$ to the Hamiltonian---where
$J(\vec x, t)$ is a given external source---triggers  a response in the expectation value of ${\cal O}$
\be \label{linear_responce} 
\langle {\cal O} (\vec x, t) \rangle_J   = -i \int_{-\infty} ^\infty \! d t' d^3 x' G_R(\vec x - \vec x \,', t-t') \, J(\vec x \,', t') + O(J^2)  
\ee
(we have assumed that the background expectation value of $\cal O$ vanishes, i.e $\langle {\cal O} (\vec x, t) \rangle_{J=0}=0$).
Its Fourier transform admits the spectral representation
\be \label{spectral}
G_R(\omega, \vec k) = \int_{-\infty}^{+\infty} \frac{d\omega_0}{\pi} \frac{i}{\omega - \omega_0 + i \epsilon} \rho(\omega_0, \vec k\, ) \; ,
\ee
where $\rho(\omega_0, \vec k\, )$---the spectral density---is a real, non-negative function (for positive $\omega_0$) that quantifies the density of states the system has at energy $\omega_0$ and momentum $\vec k$, weighed by the overlap the operator ${\cal O}$ has with them
\footnote{The finite-temperature spectral density is given by 
\begin{align}
\rho(\omega, \vec{k}) & =\sfrac{1}{2}\left(1-e^{-\beta \omega}\right)\left({\rm Tr} \,e^{-\beta H}\right)^{-1}\sum_{n,m}e^{-\beta E_n}\\
&\times (2\pi)^4\delta(\omega+E_n-E_m)\delta^3(\vec{k}+\vec{p}_n-\vec{p}_m) \, \big\vert \langle n \vert \mathcal{O} (0,\vec{0})\vert m \rangle \big\vert^2 \nonumber
\end{align}
from which the non-negativity (for positive $\omega$) follows immediately. 
}.

One is often interested in separating the real and imaginary  parts of Fourier-space correlation functions, because they contribute to different phenomena. In particular, the dissipative effects we are after will be associated with the imaginary part of $i G_R$, which, given the distributional identity
\be
\frac{1}{x + i \epsilon} = P\frac1x - i \pi \,\delta(x) \; ,
\ee
is simply the spectral density:
\be \label{GR rho}
{\rm Im} \big( i G_R(\omega, \vec k \, ) \big) = \rho(\omega, \vec k\, ) \; .
\ee

Our discussion following (\ref{FourierTransform}) thus implies that the spectral density should be infinitely differentiable for real $\omega$ and $\vec k$ at $\omega= \vec k= 0$, and that it
should admit a low-frequency, low-momentum Taylor expansion. Moreover, standard arguments (see e.g.~\cite{hartnoll}) imply that the imaginary part of $iG_R$ is odd under $\omega \to - \omega$ (while the real part is even), so that in the Taylor expansion of $\rho$ we only have odd powers of $\omega$. The dependence on $\vec k$ is constrained by rotational invariance. If ${\cal O}$ is a scalar operator, it has to involve $1$, $|\vec k|^2$,  $|\vec k|^4$, \dots; If ${\cal O}$ carries a vector index $i$, the $\vec k$-dependence of the tensor spectral density $\rho_{ij}$ will involve the combinations $\delta^{ij}$, $k^i k^j$, $|\vec k|^2 \delta^{ij}$, \dots; And so on for higher rank tensors. 
Given these properties, at very low frequencies and momenta, the spectral density of a tensor operator that transforms irreducibly under rotations can be parameterized by just {\em one} number---the first coefficient in its Taylor expansion:
\be
\rho(\omega, \vec k\, )  \simeq A \, \omega \times \delta\cdots \delta \; , \qquad \omega, k \to 0 \; ,
\ee
where $\delta\cdots\delta$ stands for the combination of Kronecker-deltas with the right symmetries 
\footnote{For any operator ${\cal O}$ of given spin $s$, there is only one possible such combination that can appear in the $\langle {\cal O} {\cal O} \rangle$ correlator. The reason is that in the tensor product of two spin $s$ representations, the singlet (spin-0) representation appears only once:
\be
(2s+1 ) \otimes (2s+1) = 1 \oplus 3 \oplus \cdots \oplus (4s+1) \; .
\ee
For instance, if ${\cal O}_{ij}$ is symmetric and traceless, that is, spin 2, its two point function at zeroth order in $\vec k$ has to take the form
\be
\langle {\cal O}_{ij} {\cal O}_{kl} \rangle \propto  \delta^{ik} \delta^{jl} + \delta^{il} \delta^{jk} - \sfrac23  \delta^{ij} \delta^{kl}  \; .
\ee}.
Notice that $A$ has to be positive, because $\rho$ is positive for positive $\omega$.

We thus see that the absence of long-range, late-time correlations in the $\chi$ sector does {\em not} forbid the existence of gapless excitations. These can exist, as long as the zero-momentum density of states {\em (i)} is a regular continuum in a neighborhood of $\omega=0$, and {\em (ii)} goes to zero at zero frequency, at least as fast as $\omega$. For instance, a $\delta$-function contribution to the spectral density,  peaked at $\omega=0$, is not allowed. This would correspond to a gapless `single particle' pole in correlators---i.e.~to an excitation with a power-law propagator at very long distances and at very late times. According to our assumptions above, this should be included in the $\phi$ sector.

\section{The actual couplings}

We can now be more specific about the structure of $S_{\rm int}$. To begin with, let us review what goes into $S_0$, which encodes the self-interactions of the $\phi$ sector. We refer the reader to refs.~\cite{DGNR, ENRW, DHNS} for the details of the derivation, and we quote here the main results only. The reader already familiar with the notation can skip directly to sect.~\ref{section S_int}.

\subsection{Review: the structure of $S_0$}
For an ordinary  fluid that does not carry conserved charges, the infrared degrees of freedom are nothing but the individual positions of its volume elements. For our purposes, it is convenient to parameterize them via three scalar fields $\phi^I(\vec x, t)$, with $I=1,2,3$, which give the comoving (or `Lagrangian') coordinates $\phi^I$ of the volume element occupying physical (or `Eulerian') position $\vec x$ at time $t$. Of course the inverse mapping $\vec x(\phi^I, t)$ offers a completely equivalent---and perhaps more intuitive---description, but we find the former more convenient because it makes implementing spacetime symmetries immediate: the $\phi^I$ behave as scalars under Poincar\'e transformations.

There are also internal symmetries acting on the $\phi^I$'s. These take a particularly simple form if we choose the comoving coordinates in the following way: for an homogeneous fluid at rest, in equilibrium at some given pressure $p_0$, we identify them with the physical ones:
\be \label{eq p0}
\langle \phi^I \rangle_{p_0}= x^I \; .
\ee
Then the internal symmetries are
\begin{align}
\phi^I & \to \phi^I + a^I   \qquad &a^I = {\rm const} \; ,  \label{shift}\\	
\phi^I & \to \phi^I + O^I {}_J \phi^J  & O^I {}_J \in SO(3)  \; ,  \label{rot} \\
\phi^I & \to \phi^I + \xi^I (\phi) &  \det \frac{\d \xi^I}{\di \phi^J} = 1 \label{diff}\; . 
\end{align}
The first and second lines encode the homogeneity and isotropy of the internal space of the fluid. The third differentiates a fluid from an isotropic solid: the dynamics of a fluid only care about compressional deformations, and are insensitive to a reshuffling of the volume elements that does not change their local density.

At lowest order in the derivative expansion, internal and spacetime symmetries force the action to take the form,
\be \label{S_0}
S_0 = \int \! d^4x \,  F(b) \; , \qquad b \equiv \sqrt{\det {\di_\mu \phi^I \di^\mu \phi^J} } \; ,
\ee
where $F$ is a generic function, which characterizes---in this language---the equation of state of the fluid (we will be more explicit about this below).

The scalar quantity $b$ defined in \eqref{S_0}, is a (relativistic) measure of the compression  level  of the fluid. Another fundamental object is the vector
\be
J^\mu \equiv \sfrac16 \epsilon^{\mu \alpha \beta \gamma}  \epsilon_{IJK} \d_\alpha \phi^I \d_\beta \phi^J \d_\gamma \phi^K \; . 
\ee
It is {\em identically} conserved, 
\be
\di_\mu J^\mu = 0
\ee
as a result of its $\epsilon$-tensor structure, and is related to $b$ and the fluid's four-velocity $u^\mu$ via
\be \label{b and u}
b^2 = -J_\mu J^\mu \; , \qquad u^\mu = \frac1b J^\mu \; . 
\ee
At lowest order in the derivative expansion, $J^\mu$ can be identified with the fluid's entropy current $s^\mu$, and $b$ with the entropy density $s$ \cite{DHNS}.

This formalism can be straightforwardly extended to accommodate a conserved charge carried by the fluid, like e.g.~baryon number \cite{DHNS, sergey}. 
One first introduces a new scalar field $\psi (\vec x, t)$ that shifts under the $U(1)$ symmetry transformation associated with the charge:
\be \label{U(1)}
\psi \to \psi + a \; ,  \qquad  a= {\rm const} \; .
\ee
Then, in order to describe an {\em ordinary} charge-carrying  fluid---as opposed to a superfluid---one promotes such a symmetry to a $\phi^I$-dependent shift symmetry:
\be \label{psi diff}
\psi \to \psi + a\big( \phi^I \big) \; .
\ee
This guarantees that the Noether current associated  with the $U(1)$  transformation \eqref{U(1)} aligns with the fluid's four velocity,
\be
j^\mu = n \, u^\mu \; ,
\ee
as befits an ordinary fluid at lowest order in the derivative expansion. The lowest order action should now read
\be \label{F(b,y)}
S_0 = \int \! d^4x \,  F(b, y)  \; ,
\ee
where $b$ is the same as above, and $y$ is defined as
\be
y \equiv u^\mu \, \d_\mu \psi \; .
\ee

The stress-energy tensor and the charge current follow straightforwardly from the action. They are
\begin{align}
T_{\mu\nu} & = \big( F_y y - F_b b \big) B^{-1}_{IJ} \, \di_\mu\phi^I \, \di_\nu \phi^J + \big(F- F_y y \big) \eta_{\mu\nu} \label{S0 Tmn}\\
j^\mu & = F_y u^\mu \; , 
\end{align}
where $F$'s subscripts denote differentiation, and $B^{-1}_{IJ}$ is the inverse of the matrix
\be
B^{IJ} \equiv \di_\mu \phi^I \di^\mu \phi^J \; . 
\ee
Equating $T^{\mu\nu}$ and $j^\mu$ above with the standard forms they take for a perfect fluid, and using thermodynamic identities, yields the following ``dictionary'' between  field theory variables and  hydrodynamical ones \cite{DHNS}:
\begin{align}
\label{basic thermo dictionary first}
\rho & = F_y y - F \\ 
p & = F- F_b b \\
s & = b \; , \qquad  T = -F_b \\  
\label{basic thermo dictionary last}
n & = F_y \; , \qquad \mu = y \; 
\end{align}
and $u^\mu$ is still given by eq.~\eqref{b and u}.
We now see that the function $F(b,y)$ is related to the equation of state: for instance, from the first, third, and fourth lines we have
\be
F(s, \mu) = n (s, \mu) \, \mu - \rho(s, \mu) \; .
\ee
The field theory language selects entropy density and chemical potential as the  natural pair of thermodynamic variables to work with.

In the following, we will consider small perturbations about an homogeneous equilibrium configuration, which,
generalizing  \eqref{eq p0} to arbitrary pressure and to the case with charge,  is described in our language by the field configuration
\be \label{bkgrd}
\phi_0^I(x) = b_0^{1/3} \, x^I \; , \qquad  \psi_0(x) = y_0 \, t \; ,
\ee
where $b_0$ and $y_0$ are the fluid's equilibrium entropy density and chemical potential. Such a configuration spontaneously breaks many of our symmetries: Lorentz boosts, completely; Spatial translations and internal $\phi^I$-shifts (eq.~\eqref{shift}), down to the diagonal combination;  Spatial and internal $\phi^I$-rotations (eq.~\eqref{rot}), down to the diagonal combination; Internal volume-preserving diffs (eq.~\eqref{diff}), completely; Time-translations and internal $\psi$-shifts (eq.~\eqref{U(1)}), down to the diagonal combination; $\phi^I$-dependent $\psi$-diffs (eq.~\eqref{psi diff}), completely. Associated with this spontaneous symmetry breaking pattern there are Goldstone excitations, which are simply fluctuations of our fields about the equilibrium configuration, 
\be 
\phi^I(x) = b_0^{1/3} \big(x^I + \pi^I \big) \; , \qquad  \psi(x) = y_0 \big(t + \pi^0 \big) \; \label{piIpi0} .
\ee

Not all the $\pi$'s feature propagating wave solutions.
Indeed, after expanding the Lagrangian \eqref{F(b,y)}  to quadratic order in fluctuations and diagonalizing it, we get \cite{Nicolis}
\be \label{L2}
{\cal L} \simeq \sfrac12 w_0 \big( \dot {\vec \pi}_L ^2 - c_s^2 ( \vec \nabla \cdot  {\vec \pi}_L )^2 \big)+ \sfrac12 w_0 \,  \dot {\vec \pi}_T ^2 
+ \sfrac12 F_{yy} y_0^2 \,  (\dot{\tilde{\pi}}^0 )^2 \; ,
\ee
where $\vec \pi_L$ and $\vec \pi_T$ are the longitudinal (curl-free) and  transverse (divergence-free) components of $\pi^I$,  and $\tilde{\pi}^0$ is a suitable linear combinations of $\pi^0$ and $\vec \pi_L$,
\be \label{pi tilde}
\dot{\tilde{\pi}}^0 = \dot \pi^0 + \sfrac{F_{by} b_0 - F_y}{F_{yy} y_0} \, \vec \nabla \cdot \vec \pi_L \; .
\ee
Moreover, $w_0 \equiv (F_y y_0 - F_b b_0)$ is the equilibrium enthalpy density $(\rho+p)_0$, $c_s^2$ is a somewhat complicated expression involving various derivatives of $F$---which corresponds precisely to the standard $\big(d p/ d \rho\big)_{S,N}$---and all derivatives of $F$ are computed at the equilibrium values $y_0$ and $b_0$.

We thus see from \eqref{L2} that only one of our Goldstones---the longitudinal part of $\pi^I$---has a standard quadratic Lagrangian for a gapless field, and wave solutions propagating at some finite speed $c_s$, with $\omega = c_s k$.
This Goldstone field corresponds to ordinary sound waves. The other Goldstones---$\vec \pi_T$ and $\tilde{\pi}^0$---do not have a gradient energy. As a result, at the order we are working, they have degenerate dispersion laws, $\omega =0 $.

For notational convenience, in the following we will not differentiate between capital $I,J, \dots$ indices and lower case $i,j,\dots$ ones. This is consistent because of the original $SO(3)_{\rm space}\times SO(3)_{\rm internal}$ rotation group, only the diagonal $SO(3)$ subgroup is preserved by the background  \eqref{bkgrd}, and under this subgroup the two sets of indices transform in the same way. In particular, the $\pi^i$ excitations transform as a vector field.

\subsection{Dissipative couplings: $S_{\rm int}$ to linear order}\label{section S_int}

We are finally in a position to write down the couplings of our $\phi$'s to the $\chi$-sector. 
There is one physical property of the $\chi$'s that we have not yet been explicit about: in a sense that we will try to make precise these degrees of freedom ``live in the fluid" simply because they ``make up'' the fluid---they are supposed to describe all the degrees of freedom of the fluid's microscopic constituents that are not explicitly taken into account by the $\phi$'s. This requirement alone should fix their transformation properties under all the symmetries that act on the $\phi$'s. 

In what follows we will restrict ourselves to the lowest order in the derivative expansion and, more importantly, to linear order in the $\pi$ fluctuations where, as we shall see shortly, the coupling to the $\chi$ sector can be read off from basic properties of Goldstone boson interactions. In order to generalize our results to higher orders in the Goldstone fields, we would need to apply systematically the so-called coset construction to our case. The coset construction allows one to write the most general interactions of Goldstone fields among themselves and with other degrees of freedom ($\chi$, in our case). One only needs to specify the symmetry breaking pattern and the transformation laws of the $\chi$ fields under the {\em unbroken} symmetries, which in our case are suitably redefined translations and rotations. Our physical requirement that the $\chi$'s ``live in the fluid'' should correspond to very specific transformation laws under such symmetries. 
The general coset construction for internal symmetries has been worked out in \cite{Weinberg_coset, CWZ, CCWZ}. It was later generalized to spontaneously broken spacetime symmetries in \cite{Volkov} (see also \cite{GHJT, APDTP} for recent applications). Our case presents yet another twist, in that certain spacetime symmetries mix with internal ones upon spontaneous breaking, in the sense that the unbroken symmetries are specific linear combinations of internal and spacetime ones. 
We leave addressing such a systematic construction to future work and content ourselves with the correct linearized description, which we discuss next.

Let us focus first on the case of a fluid without conserved charges. Suppose we start from an equilibrium configuration in which our Goldstones $\pi^I$ are set zero,
\be \label{phi0}
\phi^I_0(x) = b_0^{1/3} x^I \; .
\ee 
Then, let's turn on a small $\pi^I$ perturbation,
\be
\phi^I (x) = b_0^{1/3} \cdot \big( x^I + \pi^I(x) \big)  \; , 
\ee
with very mild spatial gradients and time-derivatives. Since $\pi^I$ appears as an addition to $x^I$, this is equivalent to performing a small spatial translation of the original equilibrium field configuration  \eqref{phi0}, weakly modulated in space and time:
\be \label{translate}
\phi^I_0\big (\vec x \big ) \to \phi^I (\vec x, t) = \phi^I_0 \big( \vec x + \vec \pi (\vec x, t) \big) \; .
\ee
We can now be precise about the meaning of ``living in the fluid'' for the $\chi$ sector: if the comoving coordinates $\phi^I$ are subjected to a weakly modulated spatial translation as in eq.~\eqref{translate}, the $\chi$ degrees of freedom undergo {\em the same} spatial  translation. But, following standard N\"other theorem-type logic, under a modulated spatial translation with parameter $\vec \pi (\vec x, t)$, the $\chi$ action changes by
\be
S_{\chi} [\chi] \to S_{\chi} [\chi] - \int \! d^4x \, \di_\mu \pi^i \, T_{\chi}^{\mu i }  \; ,   
\ee
where $T_{\chi}^{\mu i} $ is, {\em by definition}, the $\chi$ sector's contribution to the N\"other current associated with spatial translations, that is, the spatial columns of the $\chi$ sector's stress-energy tensor.
Therefore we conclude that, at linear order in $\pi^i$,
\be
\label{Sint_explicit0}
S_{\rm int} \simeq - \int \! d^4x \, \di_\mu  \pi^i  \, T^{\mu i}_{\chi} \qquad \quad \mbox{(no  charges.)}
\ee
Note that $\partial_\mu T^{\mu i}_\chi \neq 0$ and so the above interaction is non-trivial, since we are not including in $T^{\mu i}_\chi$ the $\pi$-dependent pieces that are required for conservation of the {\em total} stress-energy tensor.


A couple of comments about this expression are in order.
First,
the coupling above, while invariant under spatial translations, rotations, and $\pi$-shifts, does not seem to respect the volume-preserving  symmetry of eq.~(\ref{diff}). At linear order this symmetry requires invariance under
\be  \label{linear diff}
\vec{\pi}(t, \vec x)\to \vec{\pi}(t, \vec x) + \vec{\epsilon} \, (\vec x) \;,  \qquad \vec \nabla \cdot \vec \epsilon = 0 \; .
\ee
Since the $\vec \epsilon$ parameters are time-independent, we note that the $0$-component of eq.~(\ref{Sint_explicit0}) does respect the symmetry, whereas the spatial parts do not.  At the moment we have no satisfactory understanding of this issue, but we are confident that \eqref{Sint_explicit0} describes the correct linearized coupling of $\pi$ to the $\chi$ sector, because, as we will see in the next section, it correctly reproduces  the first-order dissipative effects of hydrodynamics.

Second,  the linear coupling of a Goldstone boson to the associated current---which we motivated via our ``living in the fluid" logic---is likely a very general feature of theories with spontaneously broken symmetries
\footnote{This is {\em not}---and has no obvious relation with---the usual statement that the current for a spontaneously broken symmetry interpolates the Goldstone particles, in the sense that given a single-Goldstone state $| \vec p \rangle$, one has
\be
\langle 0 | j^\mu | \vec p \rangle \neq 0 \; .
\ee
This interpolation property implies that the {\em full} current has terms that are linear in the Golstone field, e.g.~for relativistic theories $j_\mu = f \di_\mu \pi + \dots$ Here instead we are focusing on the terms in the current that depend on other fields---our $\chi$'s---but not on $\pi$, and we are claiming that, in the Lagrangian, the linear coupling of $\pi$ to this other sector involves precisely this $\pi$-independent part of the current.}. 
In the Appendix we show that the analog of our coupling holds for
a generic theory with a spontaneously broken internal $U(1)$ symmetry, and the logic of that example suggests that analogous results should apply for more generic (internal) symmetry breaking patterns. For spontaneously broken {\em spacetime} symmetries there will be additional subtleties, but ignoring them for the moment, we are led to postulate that in the case  of a fluid {\em with} conserved charges the leading order interaction Lagrangian will read
\be \label{Sint_explicit1}
S_{\rm int} \stackrel{?}{\simeq} - \int \! d^4x \, \big[ \di_\mu  \pi^i  \, T^{\mu i}_{\chi} + \, y_0 \, \di_\mu \pi^0 \,  j_{\chi}^\mu \big] \quad \mbox{(with charges),}
\ee
where $\pi^0$ is the Goldstone excitation of $\psi$, and the associated $y_0$ factor comes directly from its definition \eqref{piIpi0}: 
$j_{\chi}^\mu$ is the current associated with shifts of $\psi$, and turning on a $\pi^0$ field corresponds to shifting $\psi$ by $y_0 \pi^0$.
As we will see, the second term in (\ref{Sint_explicit1}) will turn out not to be the end of the story in this case---hence the question mark---but for the moment, notice that, analogously to the first term, the second term is non-trivial ($\di_\mu j_\chi^\mu \neq 0$), and {\em not} invariant under one of the symmetries, 
\be \label{linear shift} 
\pi^0(t, \vec x)\to \pi^0(t, \vec x) + a (\vec x) \; ,
\ee
which is the linearized version of \eqref{psi diff}.
We will now check that the first term in  \eqref{Sint_explicit1} reproduces correctly the first-order dissipative phenomena associated with bulk and
shear viscosity---including the celebrated Kubo relations. On the other hand, we will see that in order to model  heat conduction precisely  the second term has to be corrected, both in its overall coefficient and in its structure. We leave deriving these corrections from symmetry considerations to future work.

%

\section{Rediscovering Kubo relations}
\label{Kubo}

As advertised in sect.~\ref{idea}, we can now compute observables that involve our Goldstone excitations, and  the $\chi$ sector will contribute indirectly to these observables only via the correlators of the composite operators that couple  to our Goldstones.
Since  the only couplings that we have so far are linear in the Goldstones, the observables we are able to compute at this point have to do with the free propagation of Goldstone excitations. That is, we are able to compute the Goldstone attenuation rates. 

\subsection{Fluid without charges}
Consider first a fluid without conserved charges. Its excitation spectrum---neglecting dissipative effects---is described by the action \eqref{L2}, with the $\tilde \pi^0$ part omitted. We have a longitudinal mode $\vec \pi_L$ with $\omega = c_s k$, and two transverse modes $\vec \pi_T$ with a degenerate dispersion relation $\omega = 0$. Consider now one such excitation propagating in the fluid. Its coupling to the $\chi$ sector via the interaction \eqref{Sint_explicit0} will make it slowly decay away, eventually transferring all its initial energy to $\chi$ excitations. We can compute the rate at which this decay process takes place  at the level of the classical equations of motion for the Goldstones. We could also do the computation at the level of Feynman-diagram perturbation theory, which would be more in line with our field theoretical approach. In particular, since the attenuation rates we are after correspond to imaginary shifts in the excitations' frequencies, we should compute the $\chi$-mediated corrections to the poles of the $\pi^I$ propagator. However, as reviewed in sect.~\ref{idea}, in the in-in formalism each propagator  gets replaced by a $2\times2$ matrix
of propagators, which, at least for our simple computation, complicates unneccessarily the systematics of perturbation theory.


Following sect.~\ref{idea}, the linearized eom for $\pi^I$ derived from the Goldstone quadratic action \eqref{L2}, augmented by their interaction with the $\chi$ sector \eqref{Sint_explicit0}, is precisely what one would naively expect from  linear response theory
\footnote{We have assumed, as we did in Section \ref{idea}, that $\langle T^{\mu i}_\chi \rangle_{\pi=0}$ vanishes. 
In our formalism, the equilibrium expectation value for the fluid's {\em full} stress energy tensor is given by \eqref{S0 Tmn}, evaluated at the equilibrium configuration \eqref{bkgrd}, that is, it is fully captured by the $\phi$'s sector action $S_0$.}:
\be \label{linear eom}
w_0 \big(\omega^2 \, \pi^i - c_s^2 k^i k^j \, \pi^j \big) + i G_R^{ij}(\omega, \vec k) \, \pi^j= 0 \; ,
\ee
where $G_R^{ij}$ is the retarded two-point function of the combination that couples to $\pi^i$ in \eqref{Sint_explicit0}:
\be
G_R^{ij} (\omega, \vec k)= k_\mu k_\nu \big \langle  T_\chi^{\mu i} T_\chi^{\nu j} \big \rangle \; ,
\ee
and from now on we will use simply $\langle \dots \rangle$ to denote the Fourier transforms of {\em retarded} two-point functions, evaluated at $\omega$ and $\vec k$. Moreover, it will be understood that $G_R$ is evaluated in Fourier space, and its $\omega, \vec k$ arguments will be omitted.

In the end we are interested in the imaginary parts of the eigenfrequencies of the system, which---at leading order in perturbation theory---will be related to the imaginary part of $i G_R$. 
At this point we could parameterize the infrared behavior of ${\rm Im} \, i \cdot \big\langle T_\chi^{\mu i} T_\chi^{\nu j} \big\rangle$ as described in sect.~\ref{Low_frequency}, but, before proceeding let us massage this quantity a little in order to rewrite it in a form that the reader familiar with hydrodynamics will recognize. 
First, notice that according to \eqref{GR rho}, such a quantity is the spectral density of a composite operator ($T_\chi^{\mu i}$) in the $\chi$ sector. We argued that all local operators in the $\chi$ sector should have very well behaved spectral densities near $\omega = \vec k =0$, at least for real $\omega$ and $\vec k$, with a Taylor expansion starting as ${\rm const} \cdot \omega$, and continuing with higher powers of $\omega$ and $\vec k$. At low energies and momenta, we are interested in just that first term, which we can extract formally by taking the nested limit
\be \label{limit}
\omega \lim_{\omega \to 0}  \Big[ \sfrac1 \omega \lim_{\vec k \to 0}  \big( {\rm Im} \, i \cdot \big\langle  T_\chi^{\mu i} T_\chi^{\nu j}  \big\rangle \big) \Big] \; .
\ee
Given the  regularity of our spectral densities in the infrared, we can take the limits in any order. However, taking the limits in the order we have written them allows us to replace $T_\chi^{\mu i} $ with the {\em total}  $T^{\mu i}$,  which includes contributions coming from the Goldstone bosons. The reason is that at lowest order in the $\chi$-$\pi$ interactions and in the derivative expansion, the Golstones' contribution to {\em any} spectral density is a Dirac-delta peaked at on-shell values for $\omega $ and $\vec k$. But the limit in \eqref{limit} carefully dodges such on-shell values, both for longitudinal ($\omega = c_s k$) and for transverse ($\omega=0$) excitations. At the order we are working we thus have
\be \label{Sigma full T}
{\rm Im} \big(i G_R^{ij}  \big) \simeq \omega k_\mu k_\nu  \cdot \lim_{\omega \to 0} \sfrac1 \omega \lim_{\vec k \to 0}  
{\rm Im} \, i \cdot \big\langle T^{\mu i} T^{\nu j} \big\rangle
\ee
Then, using a standard trick---see e.g.~\cite{Jeon}---we can use conservation of the full stress-energy tensor to set to zero  terms in the correlator above that have $\mu$ or $\nu$ equal to zero. The reason is that, because of $T_{\mu\nu}$ conservation, in Fourier space we have the operator equation
\be
T^{0\alpha} = \frac{ k^k}{\omega} T^{k\alpha} \; ,
\ee
which yields zero if we take $\vec k $ to zero first, like we are doing in the limit above
\footnote{The manipulations we just performed may seem dangerous: in fact, in the last section we insisted that is important that the $\pi^i$ does {not} couple to the full $T^{\mu i}$, but only to a non-conserved part of it, so that the coupling \eqref{Sint_explicit0} is actually non-trivial.
There is no contradiction however: the divergence---or the $k^\mu$---one needs to annihilate  the full stress-energy tensor does not commute with our nested limit, so that the r.h.s.~in eq.~\eqref{Sigma full T} is actually non-zero.}.

We are thus left with
\be \label{Sigma Tij}
{\rm Im} \big(i G_R^{ij} \big) \simeq \omega k_k k_l  \cdot \lim_{\omega \to 0} \sfrac1\omega \lim_{\vec k \to 0}  
{\rm Im} \, i \cdot \big\langle T^{k i} T^{l j} \big\rangle \; .
\ee
Following sect.~\ref{Low_frequency}, we can now split the stress tensor operator $T^{ij}$ into irreducible representations of the (unbroken) rotation group, spin 0 and spin 2,
\begin{align}
T^{ij} & = T^{ij}_0 + T^{ij}_2  \\
T^{ij}_0 & = \sfrac13 \delta^{ij} \, T^{kk} \; , \qquad T^{ij}_2 = T^{ij} - \sfrac13 \delta^{ij} \,  T^{kk} \; ,
\end{align}
and parameterize the low-energy behavior of the associated spectral densities---in the nested limit we are interested in---via two free parameters $A_{0, 2}$ as
\begin{align} \label{kubo}
{\rm Im} \, i \cdot \big\langle T_0^{k i} T_0^{l j} \big\rangle & \simeq  A_0 \, \omega \cdot  \delta^{ki} \delta^{lj}  \\
{\rm Im} \, i \cdot \big\langle T_2^{k i} T_2^{l j} \big\rangle &  \simeq  A_2 \, \omega \cdot  
\big(\delta^{kl} \delta^{ij} +\delta^{kj} \delta^{il}- \sfrac23 \delta^{ki} \delta^{lj} \big) \; . \nonumber
\end{align}
We should also mention that the mixed correlator $\langle T_0 T_2\rangle$ vanishes at zero momentum, because of rotational invariance.

Plugging these parameterizations into eq.~\eqref{Sigma Tij} we get
\be
{\rm Im} \big(iG_R^{ij}  \big) \simeq \omega \, k^2 \big[  (A_0 + \sfrac43 A_2) P_L^{ij} + A_2  \, P_T^{ij}  \big] \; ,
\ee
where $P_{L,T}^{ij}$ are the longitudinal and transverse projectors
\be
P_L^{ij} = \hat k^i \hat k^j \; , \qquad P_T^{ij} = \delta^{ij}-  \hat k^i \hat k^j  \; .
\ee
The reason it's convenient to split this contribution to the $\pi^i$ eom as a sum of a longitudinal and a transverse part, is that the zeroth-order eom has a similar structure:
\be
\omega^2 \pi^i -c_s^2 k^i k^j \, \pi^j \to \big[(\omega^2 -c_s^2 k^2) P_L^{ij} + \omega^2 P_T^{ij} \big] \pi^j \;.
\ee
Then, putting everything back into eq.~\eqref{linear eom} we get immediately the imaginary parts of the (low-momentum) eigenfrequencies:
\begin{align}
\Delta \omega_L  & \simeq  - i \, \frac{ (A_0 + \sfrac43 A_2)}{2 w_0}k^2 \\
\Delta \omega_T & \simeq - i \, \frac{A_2}{w_0} \, k^2 \; .
\end{align}
These are the attenuation rates for, respectively, the longitudinal and transverse modes. 
We already see two important predictions of our  field theoretical approach.  First, the {\em dissipative} nature of the coupling \eqref{Sint_explicit0}: these imaginary frequency shifts have the right sign to make the Goldstone excitations decay in time, since the positivity of $A_{0,2}$ is guaranteed by the positivity properties of {\em any} spectral density, as reviewed in sect.~\ref{Low_frequency}.
Second, the attenuation rates scale as $k^2$ at low momenta, which agrees with the standard dissipative hydrodynamics results.

But we can go further. Comparing our attenuation rates to the standard ones in the literature---see e.g.~\cite{Jeon}---we find that our parameters $A_{0,2}$ correspond to {\em bulk and shear viscosity}, usually denoted by $\zeta$ and $\eta$:
\be \label{viscosities}
\zeta = A_0 \; , \qquad \eta = A_2 \; .
\ee
Then, our {\em definitions} of $A_{0,2}$ in eq.~\eqref{kubo}, match precisely the famous Kubo relations for bulk and shear viscosity \cite{Jeon}. This is the main result of our paper: an independent derivation of the Kubo relations via effective field theory techniques.

From now on, we will refer to the nested limit that we used above as  `the Kubo limit', and we  will denote retarded correlators in that limit by
\be
\langle \cdots \rangle _{\rm K}   \equiv \omega \lim_{\omega \to 0} \sfrac1 \omega \big( \lim_{\vec k \to 0} \langle \cdots \rangle \big) \; .
\ee

\subsection{Fluid with charges}
We can now extend the same analysis to the case of a fluid with conserved charges. However, as anticipated, the interaction Lagrangian \eqref{Sint_explicit1} now does not work equally well as for the case without charges. In particular, it  yields  the correct $\Delta \omega \sim i k^2$ scaling for the attenuation rates, but it does not reproduce the correct numerical factors in the  Kubo relations for the corresponding transport coefficients. We blame this 
on the fact that in order to guess the second term in \eqref{Sint_explicit1}, in the absence of a ``living in the fluid''-type argument we applied cavalierly  the global symmetry lesson of Appendix \ref{U(1) example} directly to our case, which involves spontaneously broken {\em space-time} symmetries, and is therefore of a slightly more subtle nature. We are confident that the coset construction will shed light on this issue. For the moment, let's see whether there exists a minimal generalization of \eqref{Sint_explicit1} that reproduces the correct physics of dissipation in this more general case. 

Let's assume that the $\chi$ sector still couples to our Goldstones only via the currents $T_{\chi}^{\mu\nu}$ and $j_\chi^\mu$. Then, by (unbroken) rotational and shift-invariance, the possible couplings at lowest order in derivatives are
\begin{align}
& \di_j \pi^i \, T_{\chi}^{j i}  \; , \quad \di_0 \pi^i \, T_{\chi}^{0 i} \; , \quad   \di_i \pi^i \, j_{\chi}^{0}  \; , \quad \di_0 \pi^i \, j_{\chi}^{i} \; ,\\
& \di_j \pi^0 \, T_{\chi}^{j 0}  \; ,  \quad \di_0 \pi^0 \, T_{\chi}^{0 0} \; , \quad \di_i \pi^0 \, j_{\chi}^{i}  \; , \quad \di_0 \pi^0 \, j_{\chi}^{0}   \; .
\end{align}
However, given what we learned above by manipulating the $\langle T_{\chi} T_{\chi} \rangle$ correlators in the Kubo limit, we notice that  correlators involving $T_{\chi}$ or $j_\chi$ with a {\em zero} index will not contribute to the imaginary parts that we are interested in. For the purposes of our computation, we can thus discard the couplings involving those composite operators. Furthermore, among  the surviving couplings, the first  is the only one involving the transverse Goldstones $\pi^i_T$: the only other possibility is $ \di_0 \pi^i \, j_{\chi}^{i}$, but recall that at lowest order in the derivative expansion  $\pi^i_T$ has  a degenerate dispersion relation $\omega = 0$, which means that, for the transverse modes, such a coupling has to be neglected at the order we are working.
By rotational invariance, the transverse modes cannot mix with either $\pi^i_L$ or $\pi^0$, and, given our success above in determining their attenuation rate in the absence of conserved charges, we want to keep their couplings to the $\chi$ sector unaltered, i.e., we want $\di_j \pi^i \, T_{\chi}^{j i}$ to appear with the same coefficient as in \eqref{Sint_explicit0}.
For the other couplings, we introduce two arbitrary coefficients. 

We thus consider the interaction Lagrangian
\be \label{Sint_explicit2}
S_{\rm int} \simeq - \int \! d^4x \, \big[ \di_j  \pi^i  \, T^{j i}_{\chi} + B  \, \di_i \pi^0 \,  j_{\chi}^i + C \,  \di_0 \pi^i \, j_{\chi}^{i} \big] \; , 
\ee
and  determine the values of $B$ and $C$ by computing the attenuation rates and matching these to known results.
We can  focus on the $\pi_L$-$\pi_0$ sector only, since for the transverse modes we have the same couplings as before, and the same analysis applies unaltered. On the other hand, when we consider the $\pi_L$ and $\pi_0$ equations of motion, we are  sensitive to new (retarded) correlators,
\be
\mbox{eom}(\pi ^\alpha) \supset iG_R^{\alpha \beta} \pi_\beta \; ,
\ee
with
\begin{align}
{\rm Im} \big(iG_R^{00}  \big) & \simeq 
B^2 \, k^ik^j \, {\rm Im} \, i \cdot \langle j^i j^{j} \rangle_{\rm K} \\
{\rm Im} \big(iG_R^{0L}  \big) & \simeq  - BC \, \omega k^i \hat k^j  \, {\rm Im} \, i \cdot \langle j^i j^{j}  \rangle_{\rm K} \\
{\rm Im} \big(iG_R^{LL} \big) & \simeq k^k k^l \, \hat k^i \hat k^j \, {\rm Im} \,  i \cdot \langle T^{ki} T^{lj}  \rangle_{\rm K}  \nonumber \\
& + C^2 \, \omega^2 \hat k^i \hat k^j    {\rm Im} \ i \cdot \langle j^i j^{j}  \rangle_{\rm K}  \; .
 \end{align}
Following the same logic as in the case without conserved charges, we have replaced the $\chi$-sector's current and stress-tensor  with the total ones.  We have also used that the mixed correlator $\langle j^i T^{jk} \rangle$ vanishes at zero momentum by rotational invariance. 

According to the general discussion of sect.~\ref{Low_frequency}, the imaginary parts of the $\langle jj \rangle_{\rm K}$ and $\langle TT \rangle_{\rm K}$ correlators have to scale as $\omega$. As we already saw, in the $TT$ case this matches the Kubo relations that determine bulk and shear viscosity---see eqs.~\eqref{kubo}, \eqref{viscosities}. In the $jj$ case, there is an analogous Kubo relation \cite{SS2}, which relates the {\em heat conductivity} $\chi$ to the coefficient of $\omega$ 
\be
 {\rm Im} \, i \cdot \langle j^i j^{j} \rangle_{\rm K} =   \chi T \, \big( \sfrac{n}{\rho+p}\big)^2 \, \omega \cdot \delta^{ij}  \; .
\ee
Putting everything together, we can rewrite the imaginary parts of the $iG_R$ entries as
\begin{align}
{\rm Im} \big(iG_R^{00}  \big) & = 
\chi T \big( \sfrac{n}{\rho+p}\big)^2 \, B^2 \, \omega k^2  \\
{\rm Im} \big(iG_R^{0L}  \big) & =  - \chi T \big( \sfrac{n}{\rho+p}\big)^2 \, BC \, \omega^2  k \\
{\rm Im} \big(iG_R^{LL} \big) & = \big( \zeta + \sfrac43 \eta \big) \, \omega k^2 + \chi T \big( \sfrac{n}{\rho+p}\big)^2 \, C^2 \, \omega^3 \; .
 \end{align}
 
To find the imaginary shifts of the eigenfrequencies, it is convenient to re-express the $iG_R$ matrix in the $\tilde \pi^0$-$\pi_L$ basis that diagonalizes the lowest-order quadratic Lagrangian \eqref{L2}. Given the definition of ${\tilde \pi}^0$, eq.~\eqref{pi tilde}, and the structure of our interaction Lagrangian, eq.~\eqref{Sint_explicit2}, this amounts to just replacing
\be
C \to C- B A \, \sfrac{k^2}{\omega^2} \; , \qquad A \equiv \sfrac{F_{by} b_0 - F_y}{F_{yy} y_0}
\ee
in the expressions for $iG_R$ above:
\begin{align}
& {\rm Im} \big(iG_R^{\tilde  0 \tilde 0}  \big)  = 
\chi T \big( \sfrac{n}{\rho+p}\big)^2 \, B^2 \, \omega k^2  \\
& {\rm Im} \big(iG_R^{\tilde 0 L}  \big)  =  - \chi T \big( \sfrac{n}{\rho+p}\big)^2 \, BC\, \omega^2  k +   \chi T \, B^2 A\, k^3  \\
& {\rm Im} \big(iG_R^{LL}  \big)  = \big( \zeta + \sfrac43 \eta \big) \, \omega k^2 + \chi T  \big( \sfrac{n}{\rho+p}\big)^2\, \big(C-BA \sfrac{k^2}{\omega^2}\big)^2 \, \omega^3 \; .
 \end{align}
 We thus get that the eigenfrequencies of the system---identified by the vanishing of the determinant of the matrix defining the eom---get shifted by
\begin{align}
\label{our heat attenuation rate}
\Delta \omega_{\tilde 0} & \simeq - i \cdot \chi T \big( \sfrac{n}{\rho+p}\big)^2 \, B^2 \, \big( \sfrac{1}{F_{yy} y_0^2} - \sfrac{A^2}{w_0 \, c_s^2} \big) \, k^2 \\
\label{our sound attenuation rate}
\Delta \omega_L & \simeq - i \cdot \Big[ (\zeta + \sfrac43 \eta) \sfrac{1}{2 w_0} + \chi T \big( \sfrac{n}{\rho+p}\big)^2 \sfrac{(c_s^2 C-BA)^2}{w_0 \, c_s^2} \Big] \, k^2 \; ,
\end{align}
where we have kept only the leading order in $k$.

We are now in a position to match our computations to the classic dissipative fluid results, which, in the case of a fluid with conserved charges, are quite messy---see e.g.~\cite{Weinberg_fluid}. At low momenta, the attenuation rates for the scalar modes are
\footnote{In fact, ref.~\cite{Weinberg_fluid} computes only the attenuation rate for the sound mode. However, following that paper's derivation
it is easy to spot another scalar mode, which corresponds to our $\tilde \pi_0$ and which we call the ``heat mode'', with an attenuation rate as given below.
}
\begin{align}
\label{heat attenuation rate}
\Delta \omega_{\rm heat} & \simeq - i \cdot \chi \, \sfrac{n \,  ({\di p}/{\di n})_T}{(\rho+p) (\di \rho/{\di T} )_n} \, k^2 \\
\label{sound attenuation rate}
\Delta \omega_{\rm sound} & \simeq - i \cdot \sfrac{1}{2(\rho+p)}\Big[ (\zeta + \sfrac43 \eta) + \chi ({\di \rho}/{\di T })_n^{-1} \\
& \times \Big( (\rho+p) - 2T  \sfrac{\di p}{\di T }\big|_n + c_s^2 T  \sfrac{\di \rho}{\di n }\big|_T - \sfrac{n}{c_s^2}  \sfrac{\di p}{\di n }\big|_T \Big) \Big] \, k^2 \nonumber
\end{align}
After a messy computation (see Appendix \ref{matching appendix}) one can see that, despite their complexity, these expressions agree with ours above if we simply choose
\be \label{BC}
B=-C= - \frac{y_0 w_0}{b_0 F_b}   \; .
\ee
That is, the two coefficients in \eqref{Sint_explicit2} have in fact the same value (up to a sign),  which in hydrodynamical/thermodynamical terms is  simply the combination ${\mu (\rho+p)}/{Ts}$. The emergence of such a simple final result  from a long series of fairly messy intermediate steps, gives us confidence in the correctness of \eqref{Sint_explicit2} with this particular choice of coefficients.
This completes our matching computation.

\section{Discussion \& Outlook}\label{sec:disc}

It is useful to summarize the salient features of our results:
From purely symmetry arguments and the principles of EFT, we were able to derive that the coupling of hydrodynamical modes to a generic thermalized sector that ``lives in fluid'' yields dissipation, with attenuation rates scaling as $k^2$. This matches well known features of dissipative effects in hydrodynamics, and is quite independent of the precise structure of the couplings that we would write down, following essentially  from the thermal nature of this extra sector.
For fluids without conserved charges, the living-in-the-fluid requirement is strong enough to determine---via symmetry considerations---the precise structure of the interactions, thus allowing us to {\it re-derive} Kubo relations. For fluids {\em with} conserved charges, we adopted a ``symmetry-inspired" ansatz for the interactions, with two free parameters, which we determined via matching a procedure.
The emergence of  a remarkably simple interaction Lagrangian, 
\be \label{Sint_final}
S_{\rm int} \simeq - \int \! d^4x \, \big[ \di_j  \pi^i  \, T^{j i}_{\chi} + \sfrac{\mu(\rho+p)}{Ts}  \big( \di_i \pi^0 -   \di_0 \pi^i \big) j_{\chi}^{i}  \, \big] \; , 
\ee
from a long series of fairly cumbersome intermediate steps, gives us confidence in the validity of the arguments behind our ansatz.

We feel that these are significant accomplishments. However, for our techniques to be a useful tool rather than simply an alternative derivation of well-know features of hydrodynamics, we need to extend our analysis beyond linear order, and, more importantly, we need to understand the systematics of the symmetry structure of the dissipative couplings. As we already emphasized, we believe that the coset construction will help us in these directions. Until then, we are left with a puzzle:
Our couplings to the $\chi$  sector do not  preserve all the symmetries we started with, in particular the  volume preserving diffs \eqref{linear diff}, and the ``modulated" shift \eqref{linear shift}. Because dissipation involves time derivatives---the factor of $\omega$ determining the infrared behavior of spectral densities---, and the symmetries above are time-independent, this was not an issue on the way to reproduce known results. However, symmetry breaking terms can be generated, for example if we compute the effects associated with the real parts of our correlators, e.g.~Re$~i\langle T^{ij}_\chi T^{kl}_\chi \rangle$. If these have frequency-independent pieces, they can yield symmetry-violating terms in the Goldstone effective action, like for instance  a  gradient energy for the transverse modes, in contradiction with standard properties of hydrodynamics. Unfortunately, from standard analytic properties of retarded Green's functions one can derive a dispersion relation of the form (see e.g.~\cite{hartnoll})
\be
\label{kkre}
{\rm Re}\big(iG_R(\omega_0,0)\big) = - \int_{-\infty}^{\infty} \frac{{\rm Im}\big(iG_R(\omega,0)\big)}{\omega-\omega_0} 
\frac{d\omega}\pi \; .
\ee
The RHS is {\em strictly} negative-definite when we take $\omega_0 \to 0$---because of the strict positivity of spectral densities---which  leads us to the result 
\be 
\lim_{\omega\to 0}~{\rm Re}\big(iG_R(\omega) \big) < 0 \ .
\ee 
That is, the unwanted frequency-independent pieces are in fact forced to be there. One could in principle  add symmetry-violating local counter-terms to cancel out the undesired, symmetry-violating contributions from the real parts of our Green's functions (these are analytic in frequency and momentum, and therefore local in position space). But this would correspond to  fine-tuning  certain Lagrangian terms to zero, which would go against the whole point of insisting on symmetries as the guiding principle to construct effective field theories: the only robust properties of physical systems should be those ensured by symmetries.
A possible resolution to this puzzle may be connected to  the validity of Eq.~\eqref{kkre}, which requires for convergence of the RHS, $\lim_{\omega \to 0} {\rm Im} \left(i G_R(\omega,0)\right) \to 0$. Even though at this stage we are not in a position to argue against such a behavior, it is nonetheless not required by any basic principle.
 
\vspace{.3cm}

We conclude with an intriguing application of our results to  black-hole physics. Ref.~\cite{GR}---from which we borrowed the techniques of sect.~\ref{idea}---analyzed dissipative effects in the dynamics of black holes, starting from a matching computation involving the absorption of gravitational waves by a black hole in isolation. A classical GR computation for such a process yields---in the language of sects.~\ref{idea} and \ref{Low_frequency}---a spectral density for the relevant composite operator scaling as $\omega$ at low frequencies \cite{GR}. Our arguments of sect.~\ref{Low_frequency} show that such a behavior is characteristic of spectral densities for local operators in thermal systems. That is, a black hole absorbs like a thermal system. This is yet another indication of the thermal nature of black holes. What is remarkable is that such an implication here follows from a purely classical computation in GR.

\vspace{.3cm}

\noindent
{\em Acknowledgements.}
During the long gestation of this paper, we have benefitted from very helpful discussions with many  of our colleagues:
Simon Caron-Huot, Sergei Dubovsky, Raphael Flauger, Walter Goldberger, Dan Green, Sean Hartnoll, Lam Hui, Guy Moore, Al  Mueller, Riccardo Penco, Eduardo Ponton, Riccardo Rattazzi, Rachel Rosen, Ira Rothstein, Slava Rychkov, Derek Teaney,  Giorgio Torrieri, and Bill Zajc. AN is particularly grateful to Sergey Sibiryakov for collaboration on the subject. His approach \cite{sergey} is in some respects complementary to ours, and we hope to see it published at some point.
The work of SE is supported by the NSF through a Graduate Research Fellowship.
The work of AN is supported by the DOE under contract DE-FG02-11ER1141743 and by NASA under contract NNX10AH14G. 
RAP is supported by the NSF under  contract AST-0807444 and by the DOE un under  contract DE-FG02-90ER40542.
The work of JW is supported by the DOE under contract  DE-FG02-92-ER40699.

\appendix

\section{Linear couplings of a $U(1)$ Goldstone}\label{U(1) example}
Consider a pair of complex scalar fields charged under a $U(1)$ global symmetry. For consistency of notation with our fluid case, let us denote them as $ \phi$ and $\chi$. Then we have a Lagrangian of the sort
\be
{\cal L}  = {\cal L}_\phi[\phi] + {\cal L}_\chi [\chi, \phi] \; ,
\ee
invariant under the $U(1)$ transformation
\be
\phi \to e^{i\alpha}\phi \, , \qquad \chi \to e^{i q \alpha}\chi
\ee
(we are allowing for different charges for $\phi$ and $\chi$.) 
Notice that for the moment we are using a slightly different notation from the main text: we are including in ${\cal L}_\chi$ both the $\chi$-sector's dynamics, and its interactions with the $\phi $ sector.
For instance, ${\cal L}_{\chi}$ might contain interactions of the form
\be \label{Lint example}
{\cal L}_{\rm \chi}\supset \lambda \big( \phi^{2q} \chi^{* \, 2} + {\rm h.c.} \big)  \; .
\ee

As in the case of spatial translations for our fluid, let us imagine now that this $U(1)$ symmetry is spontaneously broken by the vev of $\phi$: $\langle\phi\rangle = v$. 
Then, as it is standard, we can parameterize $\phi$ as
\be
\label{rhopi}
\phi = (v+\rho) e^{i\pi},
\ee
with $\pi$ being the Goldstone boson associated with the symmetry breaking, and $\rho$  the (generically) heavy radial excitation, which can be ignored at very low energies. For the sake of argument, let us thus set $\rho$ to zero from now on.
Note that the symmetry is now  realized non-linearly on $\pi$, i.e.~$\pi \to \pi+\alpha$.

We can expand the action as
\be \label{expand}
{\cal L} = {\cal L}_\phi [v \, e^{i\pi} ]  + {\cal L}_{\rm \chi}[ \chi, v \, e^{i\pi} ]
\ee
the same way we expanded $\phi^I = x^I+\pi^I$ for the fluid.  However, in this parameterization of the fields  it is not obvious that $\pi$ is derivatively coupled, as guaranteed from standard soft-pion theorems for the emission of a single soft $\pi$ quantum. For instance, from \eqref{Lint example} we get a coupling
\be
{\cal L}_{\rm \chi}\supset \lambda v^2 \big( e^{i \, 2q \pi} \chi^{* \, 2} + {\rm h.c.} \big) \; ,
\ee
which does not involve derivatives of $\pi$. This stems from a suboptimal choice of the field variables, and is easily remedied via a non-linear redefinition of the $\chi$ field, which as usual does not change the $S$-matrix:
\be
\chi = \chi' e^{i q \pi} \; .
\ee
Notice that the new $\chi'$ field is  invariant under the $U(1)$ symmetry---the transformation of $\chi$ is now carried solely by the 
$e^{iq\pi}$ factor---and the action becomes:
\be
{\cal L} =  {\cal L}_\phi [v \, e^{i\pi} ]   + {\cal L}_{ \chi}[ \chi' e^{i q \pi}, v \, e^{i\pi} ]
\ee

Let's focus on the ${\cal L}_\chi$ part.
By the $U(1)$ symmetry---which now only acts on $\pi$---this action must be invariant under constant $\pi$ shifts, $\pi \to \pi+\alpha$. Then, interpreting the $\pi$ in ${\cal L}_{ \chi}[ \chi' e^{i q \pi}, v \, e^{i\pi} ]$ as a weakly spacetime-dependent $U(1)$ transformation parameter, from N\"other's theorem we get
\be
{\cal L}_{ \chi}[ \chi' e^{i q \pi}, v \, e^{i\pi} ] = {\cal L}_{ \chi}[ \chi', v] - \di_\mu \pi \, J^\mu_\chi + \dots  \; ,
\ee
where $J^\mu_\chi$ is the $\chi$-sector's contribution to the $U(1)$ N\"other current, and we omitted terms with more $\pi$'s or more derivatives. In other words, at linear order in $\pi$ and at lowest order in the derivative expansion, the interaction between $\pi$ and the $\chi$ sector has to take the form
\be
{\cal L}_{\rm int} \simeq - \di_\mu \pi \, J^\mu_\chi \; ,
\ee
which is exactly the $U(1)$ analog of our eq.~\eqref{Sint_explicit0}. Notice that, at this order, it does not matter whether we evaluate $J^\mu_\chi$ at $\chi$ or $\chi'$, since their difference is of first order in $\pi$.
Notice also that we never really used that $\chi$ is scalar. Clearly our proof is completely general and holds for any set of charged fields $\chi$, of any spin.

As an alternative, quicker derivation of the same result, we can go back to eq.~\eqref{expand} and use the following trick (a version of St\"uckelberg's trick): We promote the global symmetry to a gauge symmetry by introducing an auxiliary gauge field ${\cal A}_\mu$ which transforms as ${\cal A}_\mu \to {\cal A}_\mu - \partial_\mu \alpha(x)$, and replace standard derivatives by covariant ones. Expanding the action to linear order in ${\cal A}_\mu$ we have
\be
\label{calLphi}
{\cal L}[\phi,\chi,{\cal A}_\mu] = {\cal L}[\phi,\chi] + J_\mu {\cal A}^\mu + {\cal O}({\cal A}^2),
\ee
where $J_\mu[\phi,\chi]$ is the conserved current (in the absence of $\cal A^\mu$) for the $U(1)$ global charge.\footnote{Note we included all the gauge-field dependence explicitly (in the ${\cal O}({\cal A}^2)$ terms), so that there is no ${\cal A}^\mu$ in $J^\mu$, which would be necessary to make it a gauge invariant expression.}
Since we promoted this symmetry to a gauge transformation we are guaranteed that $\pi$ {\it disappears} from the action, because it can be absorbed into ${\cal A}_\mu$ by choosing $\alpha = - \pi$. This means that, to linear order in $\pi$, we wind up with the coupling
\footnote{Notice that $J_\mu$ in Eq. (\ref{calLphi}) does depend on $\pi$, e.g.~$J_\mu \supset f \partial_\mu \pi$, for some `decay constant' $f$. However this dependence, and the one stemming from  ${\cal L}[\phi,\chi]$,  is ultimately absorbed into ${\cal A}^\mu$ including the ${\cal O}({\cal A}^2)$ piece. That is why only $J_\chi^\mu$ enters in the coupling to linear order. 
}
\be
J_\chi^\mu \partial_\mu \pi,
\ee
where $J_\chi^\mu$ is the $\chi$-dependent component of the full current at zeroth order in $\pi$.\footnote{The alert reader may have already recognized this is the way longitudinal gauge bosons couple to matter.} Hence we conclude that, at leading order in the perturbations, the Goldstone boson couples to the $\pi$-independent part of the current associated with the broken symmetry.\\

The introduction of ${\cal A}_\mu$ is equivalent to working in the so called {\it unitary} gauge, where the Goldstones are set to zero and their interactions are encoded in the gauge field. The previous analysis suggests that one could perform similar manipulations in the case of our fluid, where the Goldstone fields $\pi^I$ are associated with the breaking of spatial translations. Now, to go to the unitary gauge we must introduce the gauge field associated with spatial translations, namely the metric perturbation $h_{\mu I}$, and the broken generators are the $T^{\mu I}$ components of the stress energy tensor. Hence, the coupling must read: $T^{\mu I}_\chi h_{\mu I}$. To introduce the pions we do as before, which in our case entails schematically
\be
h_{\mu I} = \partial_{(\mu} \alpha_{I)} \to \partial_{(\mu} \pi_{I)}.
\ee
This viewpoint might prove useful in extending our results to non-linear order.


\section{Matching the attenuation rates}\label{matching appendix} 

In order to match our computed attenuation rates for the sound and heat modes given by (\ref{our heat attenuation rate}) and (\ref{our sound attenuation rate}) with the known values given by (\ref{heat attenuation rate}) and (\ref{sound attenuation rate}) and fix our free parameters $B$ and $C$ we must express many thermodynamic quantities such as $c_s^2$, $(\di \rho/\di T)_n$, etc. in terms of our effective theoretical variables $b, y$ and various derivatives of $F$.  Explicitly they are as follows: 
\begin{align}
&c_s^2 =\frac{{\rm d}(F-F_b b_0)}{{\rm d} (F_y y_0-F)}\Big \vert_{{\rm d}(F_y/b)=0}=\frac{(F_y-F_{by}b_0)^2-b_0^2 F_{yy}F_{bb}}{w_0 F_{yy}},\\
&\frac{\di \rho}{\di T}\Big \vert_n = \frac{d (-F+F_y y_0)}{d(-F_b)}\Big \vert_{d F_y=0}=-\frac{F_{yy}F_b}{(F_{by}^2-F_{bb}F_{yy})},\\
&\frac{\di p}{\di T}\Big \vert_n =\frac{d (F-F_b b_0)}{d (-F_b)}\Big \vert_{d F_y=0}=b_0+\frac{F_y F_{yb}}{F_{bb}F_{yy}-F_{by}^2},\\
&\frac{\di p}{\di n}\Big \vert_T =\frac{d (F-F_b b_0)}{d(F_y)}\Big \vert_{d F_b=0}=-\frac{F_y F_{bb}}{F_{by}^2-F_{bb}F_{yy}} \; .
\end{align}
(For $c_s^2$, the $F_y/b = n/s = {\rm const}$ constraint is equivalent to the usual one, $S = {\rm const}$, $N={\rm const}$ \cite{Nicolis}.)
Along with the relations (\ref{basic thermo dictionary first})-(\ref{basic thermo dictionary last}) this is all we need to perform the matching. We find that for the two computations to coincide 
\be
B=-C\quad \text{with } \left| B \right|=-\frac{y_0 w_0}{b_0 F_b}=\frac{\mu (\rho+p)}{T s}
\ee
as quoted in the text.


\begin{thebibliography}{99}


\bibitem{DGNR}
  S.~Dubovsky, T.~Gregoire, A.~Nicolis and R.~Rattazzi,
  ``Null energy condition and superluminal propagation,''
  JHEP {\bf 0603}, 025 (2006)
  [arXiv:hep-th/0512260].

\bibitem{ENRW}
  S.~Endlich, A.~Nicolis, R.~Rattazzi and J.~Wang,
``The quantum mechanics of perfect fluids,''
  JHEP {\bf 1104}, 102 (2011)
  [arXiv:1011.6396 [hep-th]].

\bibitem{NS}
A.~Nicolis and D.~T.~Son,
``Hall Viscosity from Effective Field Theory,''
arXiv:1103.2137 [hep-th].

\bibitem{DHNS}
S.~Dubovsky, L.~Hui, A.~Nicolis and D.~T.~Son,
``Effective Field Theory for Hydrodynamics: Thermodynamics, and the Derivative Expansion,''
arXiv:1107.0731 [hep-th].

\bibitem{DHN}
S.~Dubovsky, L.~Hui and A.~Nicolis,
``Effective Field Theory for Hydrodynamics: Wess-Zumino Term and Anomalies in Two Spacetime Dimensions,''
arXiv:1107.0732 [hep-th].

\bibitem{Nicolis}
A.~Nicolis,
``Low-Energy Effective Field Theory for Finite-Temperature Relativistic Superfluids,''
arXiv:1108.2513 [hep-th].

\bibitem{SSu} 
  D.~T.~Son and P.~Surowka,
  ``Hydrodynamics with Triangle Anomalies,''
  Phys.\ Rev.\ Lett.\  {\bf 103}, 191601 (2009)
  [arXiv:0906.5044 [hep-th]].
  
\bibitem{Jeon}
S.~Jeon,
``Hydrodynamic Transport Coefficients in Relativistic Scalar Field Theory,''
Phys.\ Rev.\ D {\bf 52} (1995) 3591
[arXiv:hep-ph/9409250].

\bibitem{SS1}
D.~T.~Son and A.~O.~Starinets,
``Viscosity, Black Holes, and Quantum Field Theory,''
Ann.\ Rev.\ Nucl.\ Part.\ Sci.\ {\bf 57} (2007) 95
[arXiv:0704.0240 [hep-th]].

\bibitem{DG} 
  P.~Danielewicz and M.~Gyulassy,
  ``Dissipative Phenomena in Quark Gluon Plasmas,''
  Phys.\ Rev.\ D {\bf 31}, 53 (1985).

\bibitem{KSS}
P.~Kovtun, D.~T.~Son and A.~O.~Starinets,
``Viscosity in Strongly Interacting Quantum Field Theories from Black Hole Physics,''
Phys.\ Rev.\ Lett.\ {\bf 94} (2005) 111601
[arXiv:hep-th/0405231].

\bibitem{Cremonini}
S.~Cremonini,
``The Shear Viscosity to Entropy Ratio: a Status Report,''
Mod.\ Phys.\ Lett.\ B {\bf 25} (2011) 1867
[arXiv:1108.0677 [hep-th]].

\bibitem{etascaus}
  M.~Brigante, H.~Liu, R.~C.~Myers, S.~Shenker and S.~Yaida,
  ``The Viscosity Bound and Causality Violation,''
  Phys.\ Rev.\ Lett.\  {\bf 100} (2008) 191601
  [arXiv:0802.3318 [hep-th]].

\bibitem{adams} A.~Adams, N.~Arkani-Hamed, S.~Dubovsky, A.~Nicolis and R.~Rattazzi,
  ``Causality, analyticity and an IR obstruction to UV completion,''
  JHEP {\bf 0610}, 014 (2006)
  [hep-th/0602178].

\bibitem{wwscat} J.~Distler, B.~Grinstein, R.~A.~Porto and I.~Z.~Rothstein,
  ``Falsifying Models of New Physics via WW Scattering,''
  Phys.\ Rev.\ Lett.\  {\bf 98}, 041601 (2007)
  [hep-ph/0604255].

\bibitem{GR} 
  W.~D.~Goldberger and I.~Z.~Rothstein,
  ``Dissipative effects in the worldline approach to black hole dynamics,''
  Phys.\ Rev.\ D {\bf 73}, 104030 (2006)
  [hep-th/0511133].

\bibitem{dispin} R.~A.~Porto,
  ``Absorption effects due to spin in the worldline approach to black hole dynamics,''
  Phys.\ Rev.\ D {\bf 77}, 064026 (2008)
  [arXiv:0710.5150 [hep-th]].

\bibitem{diseft}  D.~Lopez Nacir, R.~A.~Porto, L.~Senatore and M.~Zaldarriaga,
 ``Dissipative effects in the Effective Field Theory of Inflation,''
  JHEP {\bf 1201}, 075 (2012)
  [arXiv:1109.4192 [hep-th]].
  
\bibitem{diseft2}  D.~Lopez Nacir, R.~A.~Porto and M.~Zaldarriaga,
  ``The consistency condition for the three-point function in dissipative single-clock inflation,''
  JCAP {\bf 1209}, 004 (2012)
  [arXiv:1206.7083 [hep-th]].

\bibitem{galley} C.~R.~Galley,
  ``The classical mechanics of non-conservative systems,''
  arXiv:1210.2745 [gr-qc].

\bibitem{MSS} 
  P.~Meade, N.~Seiberg and D.~Shih,
  ``General Gauge Mediation,''
  Prog.\ Theor.\ Phys.\ Suppl.\  {\bf 177}, 143 (2009)
  [arXiv:0801.3278 [hep-ph]].

\bibitem{Goldstein}
H.~Goldstein, C.~Poole and J.~Safko,
``Classical mechanics,''
{\it San Francisco: Addison Wesley (2002) 638 p.}


\bibitem{Jordan} 
  R.~D.~Jordan,
  ``Effective Field Equations for Expectation Values,''
  Phys.\ Rev.\ D {\bf 33}, 444 (1986).

\bibitem{Maciejko}
J.~Maciejko,
``An Introduction to Nonequilibrium Many-Body Theory," 
available for download at 
\url{http://www.physics.arizona.edu/~stafford/Courses/560A/nonequilibrium.pdf}

\bibitem{CH} 
  E.~Calzetta and B.~L.~Hu,
  ``Nonequilibrium Quantum Fields: Closed Time Path Effective Action, Wigner Function and Boltzmann Equation,''
  Phys.\ Rev.\ D {\bf 37}, 2878 (1988).

\bibitem{hartnoll}
  S.~A.~Hartnoll,
  ``Lectures on holographic methods for condensed matter physics,''
  Class.\ Quant.\ Grav.\  {\bf 26}, 224002 (2009)
  [arXiv:0903.3246 [hep-th]].
  
\bibitem{sergey}
S. Sibiryakov, unpublished.
  
\bibitem{Weinberg_coset} 
  S.~Weinberg,
  ``Nonlinear realizations of chiral symmetry,''
  Phys.\ Rev.\  {\bf 166}, 1568 (1968).

\bibitem{CWZ} 
  S.~R.~Coleman, J.~Wess and B.~Zumino,
  ``Structure of phenomenological Lagrangians. 1.,''
  Phys.\ Rev.\  {\bf 177}, 2239 (1969).

\bibitem{CCWZ} 
  C.~G.~Callan, Jr., S.~R.~Coleman, J.~Wess and B.~Zumino,
  ``Structure of phenomenological Lagrangians. 2.,''
  Phys.\ Rev.\  {\bf 177}, 2247 (1969).

\bibitem{Volkov} 
  D.~V.~Volkov,
  ``Phenomenological Lagrangians,''
  Fiz.\ Elem.\ Chast.\ Atom.\ Yadra {\bf 4}, 3 (1973).

\bibitem{APDTP} 
  C.~Armendariz-Picon, A.~Diez-Tejedor and R.~Penco,
  ``Effective Theory Approach to the Spontaneous Breakdown of Lorentz Invariance,''
  JHEP {\bf 1010}, 079 (2010)
  [arXiv:1004.5596 [hep-ph]].

\bibitem{GHJT} 
  G.~Goon, K.~Hinterbichler, A.~Joyce and M.~Trodden,
  ``Galileons as Wess-Zumino Terms,''
  arXiv:1203.3191 [hep-th].

\bibitem{SS2} 
  D.~T.~Son and A.~O.~Starinets,
  ``Hydrodynamics of r-charged black holes,''
  JHEP {\bf 0603}, 052 (2006)
  [hep-th/0601157].

\bibitem{Weinberg_fluid} 
  S.~Weinberg,
  ``Entropy generation and the survival of protogalaxies in an expanding universe,''
  Astrophys.\ J.\  {\bf 168}, 175 (1971).

%


\end{thebibliography}
\end{document}